\documentclass[article]{mn2e}
\usepackage{epsfig,rotating}
\usepackage{mnras_cite}

\newcommand{\etal}{et~al.\ }

\newcommand{\ginga}{{\it Ginga}}
\newcommand{\cgro}{{\it CGRO}}
\newcommand{\sax}{{\it BeppoSAX}}
\newcommand{\rxte}{{\it RXTE}}

\newcommand{\msun}{\thinspace\hbox{$M_{\odot}$}}
\newcommand{\mdot} {\thinspace\hbox{$\dot M$}}

\def\spose#1{\hbox to 0pt{#1\hss}}
\def\lta{\mathrel{\spose{\lower 3pt\hbox{$\mathchar"218$}}
     \raise 2.0pt\hbox{$\mathchar"13C$}}}
\def\gta{\mathrel{\spose{\lower 3pt\hbox{$\mathchar"218$}}
     \raise 2.0pt\hbox{$\mathchar"13E$}}}

\title
[Long-term X-ray variability and state transition of GX\,339--4]
{Long-term X-ray variability and state transition of GX\,339--4}

\author[A. K. H. Kong, P. A. Charles, E. Kuulkers and S. Kitamoto]
{A. K. H. Kong$^1$\thanks{Present address: Harvard-Smithsonian Center
for Astrophysics, 60 Garden Street, Cambridge, MA 02138, USA; Email: 
akong@cfa.harvard.edu},
P. A. Charles$^{1,2}$, E. Kuulkers$^{3,4}$ and S. Kitamoto$^5$\\
$^1$ Department of Astrophysics, University of Oxford, Keble Road, Oxford OX1 
3RH\\
$^2$ Department of Physics and Astronomy, University of Southampton, 
Southampton SO17 1BJ\\
$^3$ Space Research Organization Netherlands, Sorbonnelaan 2, 3584 CA Utrecht, 
the Netherlands \\
$^4$ Astronomical Institute, Utrecht University, P.O. Box 80000, 3507 TA
Utrecht, the Netherlands \\
$^5$ Department of Physics, College of Science, Rikkyo University,
3-34-1, Nishi-Ikebukuro, Toshima-ku, Tokyo 171-8501, Japan\\
}
\date{Accepted. Received.}
\begin{document}
\maketitle

\begin{abstract}
    
  With extensive monitoring data spanning over 30 years from {\it Vela
    5B}, {\it Ariel 5}, {\it Ginga}, {\it Compton Gamma Ray
    Observatory}, {\it Rossi X-ray Timing Explorer} and \sax, we find
  evidence for long-term X-ray variability on timescales
  $\sim$~190--240~d from the black hole low-mass X-ray binary system
  GX\,339--4. Such variability resembles the outburst cycle of Z
  Cam-type dwarf novae, in which the standard disc instability model
  plays a crucial role. If such a model is applicable to GX\,339--4,
  then the observed variability might be due to the irradiation of an
  unstable accretion disc. We show that within the framework of the
  X-ray irradiation model, when the accretion rate exceeds a critical
  value, GX\,339--4 enters a `flat-topped' high/soft state, such as
  seen in 1998, which we suggest corresponds to the `standstill' state
  of Z Cam systems.

\end{abstract}

\begin{keywords}
accretion, accretion discs -- binaries: close -- black hole physics -- stars: 
individual (GX\,339--4) -- X-rays: stars
\end{keywords}

\section{Introduction}
\label{sec:intro}

The black hole candidate GX\,339--4 was discovered by
\scite{markert73} with the {\it OSO-7} satellite and was soon noted
for its similarity in X-ray behaviour to the classical black hole
candidate Cyg\,X--1 (\pcite{markert73}; \pcite{maejima84};
\pcite{dolan87}). The source exhibits aperiodic and quasi-periodic
modulations on time scales spanning milliseconds to years and over a
wide range of wavelengths. Unlike typical soft X-ray transients (SXTs)
which are barely detected in their quiescent state, GX\,339--4 spends
most of the time in the so-called X-ray low/hard state (LS) like a
faint persistent source. During the LS, the energy spectrum can be
described as a 
power-law with photon index of $\sim$~1.5--2 (see Appendix for full
list of references). It changes to the
high/soft state (HS) occasionally (e.g. \pcite{maejima84,belloni99})
which are similar to the outbursts of transient sources. During the
HS, it becomes brighter (in the 2--10 keV band) by a factor of
$\sim$~5--100 (see e.g. \pcite{kong00}, and references therein) and
exhibits an ultra-soft spectral component plus a steeper power-law
component. The `low' and `high' states referred to here are also
called the `hard' and `soft' states so as to reflect the spectral
behaviour. However, the luminosity of the `low' state can sometimes be
higher than the `high' state (see Appendix) and hence it can be
confusing to define the `state' solely by the intensity of the X-rays.
Therefore, a knowledge of the X-ray spectra are
essential to distinguish different states accurately (see Table 1 of
\pcite{kong00} for the spectral and temporal properties of different
states).

In addition to the LS and HS, GX\,339--4 exhibits a very high state (VHS;
\pcite{miyamoto91}) with a higher X-ray luminosity than in the HS (by
a factor of $\sim$~3). Note that the VHS also occurs in several other black
hole soft X-ray transients (BHSXTs): GS\,1124--683 \cite{ebisawa94},
XTE\,1550--564 \cite{sobczak99} and XTE\,J1748--288 \cite{revnivtsev00}. 
An intermediate state (IS) in e.g. GX\,339--4 \cite{mendez97} and 4U\,1630--47
\cite{dieters00} has also been reported  and
its spectral and timing properties are similar to the VHS but with a
much lower luminosity. Finally, every now and then GX\,339--4 enters
an `off' state (see \pcite{markert73}; \pcite{motch85}; \pcite{ilovaisky86};
\pcite{asai98}; \pcite{kong00}), in which the X-ray fast time
variability and spectral shape are consistent
with that seen in the LS (\pcite{mendez97}), except that the 2--10~keV flux
is at least $\sim 10$ times lower than in the LS. It was concluded
that the `off' state is indeed an extension of the LS, but with lower
luminosity (\pcite{kong00}; \pcite{corbel00}). It is worth noting
that GX\,339--4 and GS\,1124--683 \cite{ebisawa94} are the only X-ray
sources observed in all these states; however, only GX\,339--4 visits
the different states so frequently (except for the VHS). We show in
Figures 1--3 the different states of GX\,339--4 in the past 30 years.

X-ray variability on long timescales (from days to years) has been
found in many low-mass and high-mass X-ray binaries, but its origin is still an open question
(e.g.
\pcite{priedhorsky87,schwarzenberg92,wijers99,ogilvie00}). Long-term
X-ray variability has also been seen in the persistent black hole
candidates Cyg\,X--1 (e.g. \pcite{kitamoto00}) and LMC\,X--3
(e.g. \pcite{paul00}). The timescales of their variability range from
100 days to 300 days. Given the similarity in X-ray properties between
GX\,339--4 and Cyg\,X--1, it is intriguing to investigate the
long-term X-ray variability of GX\,339--4. We therefore exploit the
long-term capabilities of the all sky monitoring
instruments on board the {\it Vela 5B}, {\it Ariel 5}, \ginga, {\it
Compton Gamma Ray Observatory} and {\it Rossi X-ray Timing Explorer} to study 
the X-ray behaviour of this source on timescales of months to years.

\section{Long-term X-ray Observations}

GX\,339--4 has been monitored by several X-ray missions in the last 30
years. We describe below all the instruments which contribute to the
long-term X-ray light curves for this work. In addition, a new \sax\ 
observation of the source during the `off' state is reported so that
we can have a better understanding of the current status of the
source.  GX\,339--4 has also been studied by other pointed
observations which are summarised in the Appendix.

\subsection{{\it Vela 5B}}
The {\it Vela 5B} satellite \cite{conner69} monitored the X-ray sky
from 1969 August to 1976 August (MJD 40367--43937) in two energy channels, 3--12 keV and
6--12 keV, the archival results of which are available from the High
Energy Astrophysics Science Archive Research Center
(HEASARC)\footnote{http://heasarc.gsfc.nasa.gov}.  The light curve
(see Fig.~\ref{fig:av}) is made up from 508 data points using only the
first channel (3--12 keV) since it has higher signal-to-noise ratio
compared to the second channel (6--12~keV). Due to its limited
temporal resolution and sensitivity, the data points exhibit a large
scatter and only some outburst-like events were detected.
It is interesting to note that there is a HS at $\sim$~MJD~41370 and
41660 (as indicated in Fig.~\ref{fig:av}) according to the pointed
observations made by satellite {\it OSO-7} (\pcite{markert73}, see
Appendix); however the intensity is not that high comparing to other
data points.

\subsection{{\it Ariel 5}}
\begin{figure}
\begin{center}
{\rotatebox{0}{\psfig{file=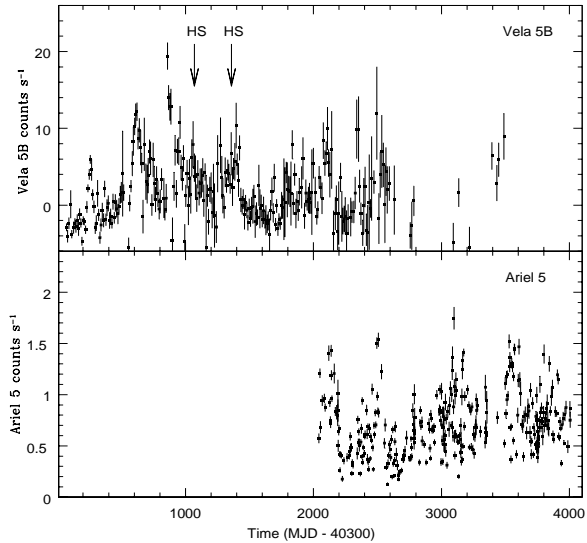,height=7.8cm,width=8.3cm}}}
\caption{X-ray light curves of GX\,339-4 as observed by: (upper panel)
{\it Vela 5B} from 1969 -- 1976. Arrows indicate the HS observed with {\it OSO-7}; (lower panel) {\it Ariel 5} ASM from 1974 -- 1980.  Both datasets were rebinned by a factor for 4 for clarity.}
\label{fig:av}
\end{center}
\end{figure}

The {\it Ariel 5} ASM experiment \cite{holt76} consisted of a pair of
X-ray pinhole cameras with position sensitive proportional counters
(3--6 keV) that covered 75\% of the sky during each orbit
($\sim$\,100\,mins). From the archival HEASARC {\it Ariel 5} database,
we obtained 2605 data points on GX\,339--4 spanning the period from
1974 October to 1980 March (MJD 42338--44308; Fig.~\ref{fig:av}). Note that part of the
{\it Ariel 5} data was collected simultaneously with the {\it Vela 5B}
(see Fig.~\ref{fig:av}) and the count rate variations are consistent
with each other. As for the {\it Vela 5B} data, the light curve shows
considerable scatter as well as some clear flaring episodes. No pointed
observations were carried out in this period and hence the X-ray
`state' of the source is not clear.

\subsection{\ginga\ ASM}
\label{sec:ginga}
\begin{figure}
\begin{center}
{\rotatebox{0}{\psfig{file=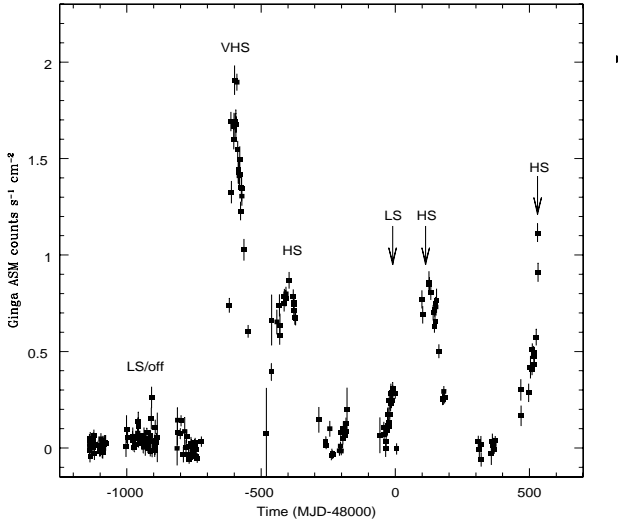,height=7.5cm,width=8.3cm}}}
\caption{{\it Ginga} ASM 1--20~keV light curve of GX\,339--4.
This clearly shows that the source has gone through several
different state transitions between 1987 and 1990. The arrows mark the
time of {\it Granat} observations (see Appendix), which provide
additional confirmation of the X-ray `state'.}
\label{fig:ginga}
\end{center}
\end{figure}

The \ginga\ ASM \cite{tsunemi89} monitored the X-ray sky in the 1--20~keV band 
from 1987 February to
1991 October. The effective area of the \ginga\ ASM was about
420~cm$^{2}$ with a $45^{\circ}\times 1^{\circ}$ field-of-view. The
sky-scanning observations were performed at intervals of a few days
and covered about 70\% of the sky. The typical exposure time for each
scan across each observed source is about 3--18~s.  During the
observations from 1987 March to 1991 October (MJD 46860--48531), 215 data points were
collected and it is very clear that four state transitions are seen in the
light curve (Fig.~\ref{fig:ginga}), indicating excursions from the LS
to the VHS/HS. The LS, HS and VHS state identifications were confirmed by
the energy and power spectra obtained with the \ginga\ LAC
\cite{miyamoto91}, {\it Granat} \cite{grebenev93} and BATSE
\cite{harmon94} instruments. During the LS, the corresponding flux is always below
100~mCrab (1--6~keV) and in 1988 August ($\sim$ MJD 47374), the source flux increased
dramatically up to 1~Crab in $\sim$~10 days. Subsequent \ginga\ LAC
observations in 1988 September ($\sim$ MJD 47405) revealed that the source was in a
VHS. The flux then decreased to about 500~mCrab over $\sim$~55 days,
which presumably returned to the HS. By the end of 1989, the source
again went to the LS until the other LS/HS transition between 1990
August and December ($\sim$ MJD 48104--48226). The last LS/HS transition observed by \ginga\ took place in 1991
August which was observed simultaneously with BATSE and will be discussed in 
\S\ref{sec:batse}.

\subsection{\rxte\ ASM}
\label{sec:339rxte}

\begin{figure*}
\begin{center}
{\rotatebox{0}{\psfig{file=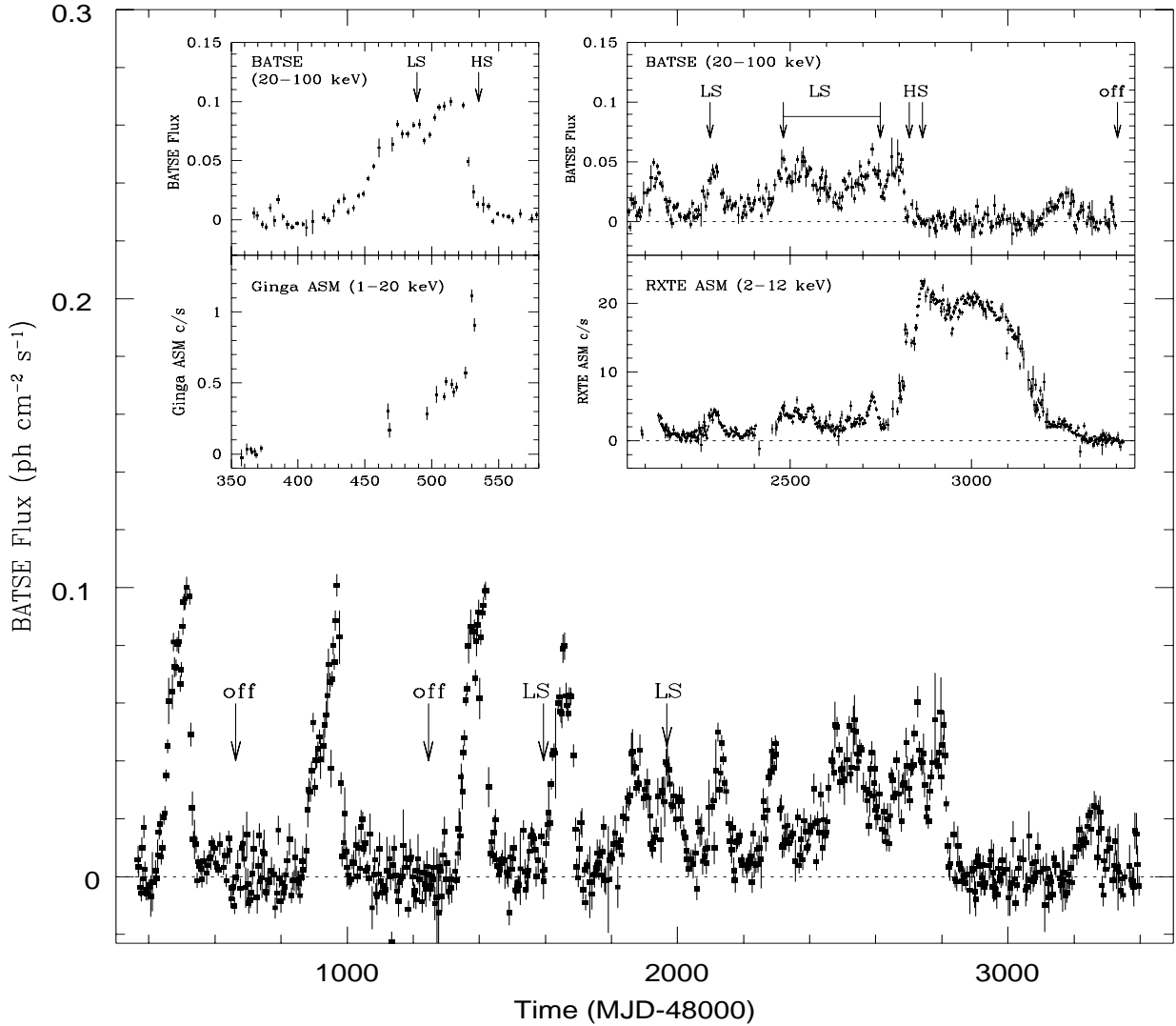,height=15cm,width=17.6cm}}}
\caption{BATSE 20--100~keV light curve of GX\,339--4 from 1991 --
1999. The data were rebinned by a factor of 2 for clarity. Also shown 
in the insets are the simultaneous BATSE and
\ginga\ ASM light curves (left) and simultaneous BATSE and
\rxte\ ASM light curves (right). The \rxte\ ASM data were
rebinned by a factor of 2. Note that during a LS/HS
transition, the BATSE flux drops to below the detection limit, while
the soft X-rays (\ginga\ or \rxte) increase dramatically. The arrows
mark the time of pointed observations (see Appendix) and the states
are defined by the associated observations.} 
\label{fig:batse}
\end{center}
\end{figure*}

The All Sky Monitor (ASM; \pcite{levine96}) on board the {\it Rossi X-ray 
Timing Explorer} (\rxte; \pcite{bradt93}) has
monitored GX\,339--4 several times daily in its 2--12 keV pass-band
since 1996 February.
The source remained at a low flux level ($\sim 2$~ASM cts/s or 27~mCrab) 
until early 1998 January ($\sim$ MJD 50810) although some variations
were seen (see inset of Fig.~\ref{fig:batse}). Extensive pointed Proportional 
Counter Array (PCA)
observations during this period indicate that it was in the LS
\cite{wilms99,belloni99}. After MJD 50800, the source flux increased 
dramatically
to $\sim 20$ ASM cts/s (270~mCrab) where it stayed for about 200 days before 
declining. \scite{belloni99} reported that the source underwent a LS to HS 
transition, probably
through an IS ($\sim$ MJD 50820). The source changed back
to the LS again in 1999 February ($\sim$ MJD 51200).
After 1999 June ($\sim$ MJD 51330), the ASM count rate dropped further and the
source intensity fell below the 3-$\sigma$ detection level. \scite{kong00} 
found that the source entered the `off' state at that time from \sax\ and 
optical observations.
Note that the ASM count rate is also correlated with the hard X-rays
as observed by BATSE (see next subsection). Interestingly, radio
emission is also correlated with the X-ray during
these different states \cite{fender99,corbel00}.

\subsection{\cgro\ BATSE}
\label{sec:batse}
The Burst and Transient Source Experiment (BATSE) on board the {\it 
Compton Gamma Ray Observatory (CGRO)} was operated continuously 
from April 5, 1991 (MJD 48351) until its re-entry on June 4, 2000 (MJD
51699).  BATSE
consists of eight identically configured detector modules with energy 
channels spanning from 20 to 600 keV (see \pcite{fishman89}).
The GX\,339--4 observations presented here were taken from the archival
database which consists of data from 1991 April to 1999 August (MJD
48370--50554) in the
20--100~keV band. In constructing the light curve, the detector
count rate is obtained by the Earth occultation technique \cite{harmon94}
and is fitted by an optically thin thermal bremsstrahlung (OTTB) model at a fixed
temperature $kT=60$~keV \cite{rubin98}, resulting in the photon fluxes
presented here. The data presented in Fig.\ref{fig:batse} were rebinned by a 
factor of 2.

The BATSE light curve (see Fig.~\ref{fig:batse}) shows about ten hard
X-ray outbursts.  The first hard X-ray outburst in 1991 August ($\sim$
MJD~48470) was
accompanied by \ginga\ ASM observations and two pointed observations
by {\it Granat} (inset of Fig.~\ref{fig:batse}). The soft and hard
X-rays are more or less correlated until the hard X-rays reached
maximum. The hard X-rays then decreased sharply and the source became
very soft with a much steeper power law spectrum
\cite{grebenev93,harmon94}; the source changed from the LS (as
observed with the \ginga\ ASM) to the HS. The HS lasted for about 70
days, during which near-simultaneous radio data were available
\cite{corbel00} and then the source entered a possible `off' state due
to the non-detection of soft X-ray emission in the 3--10 keV band
\cite{grebenev93}.  The source then underwent three major hard X-ray
outbursts before 1996 (MJD 50083) and a detailed energy spectral
analysis of the BATSE data was presented by \scite{harmon94} and
\scite{rubin98}. In particular, the second one (MJD~49340--49420) is
similar to the very first outburst (observed by  {\it Ginga} observations), and
radio observations \cite{corbel00} suggest that the duration of the HS
(after the sharp decrease in flux at $\sim$~MJD~49420) is also about
70 days.  It is not clear whether there is a HS occurring between
MJD~49000--49300 and MJD~49500--50100 since there is no soft X-ray
and/or pointed data, but radio observations after
MJD~49500 indicate that the source stayed in the LS \cite{corbel00}.

After the launch of \rxte,
the source was monitored regularly by both the \rxte/ASM and BATSE and it is
easily seen in the simultaneous light curves that the soft and hard
X-rays are (anti) correlated in their gross behaviour, as already suggested from the
simultaneous \ginga\ ASM and BATSE light curves. In addition, some
`mini'-outbursts with shorter recurrence times were also seen in the
BATSE light curve. Extensive pointed \rxte\ observations suggest that
the source was in the LS during these periods \cite{wilms99,belloni99}.
BATSE initially observed three strong outbursts
with near equal intensities and separations ($\sim 400$ days), the
subsequent outbursts clearly show that the time between outbursts was
variable, with a significant correlation between outburst fluence and
time since the last outburst \cite{rubin98}. The source changed from the LS to 
HS in 1998
January ($\sim$ MJD~50814) according to \rxte\ (ASM and PCA) observations \cite{belloni99} and 
the BATSE
flux dropped to below significant detection levels. Such a LS/HS transition
was also accompanied by non-detection of radio emission
\cite{corbel00}. The HS ended in 1999 February ($\sim$ MJD~51210) and the return to the LS was 
accompanied by the reappearance of the hard X-rays and radio emission 
\cite{corbel00}. Following the LS, the source entered the `off' state (see 
\S\ref{sec:339rxte}) with a small drop in the BATSE flux.

\subsection{\sax}
\label{sec:sax}

We observed GX\,339--4 with the Narrow Field Instruments (NFI) on board {\it
BeppoSAX} between March 23.9 and 24.6, 2000 UT. The NFI consist of two
co-aligned imaging instruments providing 
a field of view of $37' \times 57'$: the
Low-Energy Concentrator Spectrometer (LECS; 0.1--10 keV; Parmar et al.
1997a) and the Medium Energy Concentrator Spectrometer (MECS; 1.6--10.5 keV;  
Boella et al. 1997).  The other two NFI, non-imaging instruments are
the Phoswich Detector System (PDS; 12--300keV; Frontera et al. 1997) and
the High-Pressure Gas Scintillation Proportional Counter (HP--GSPC; 4--120 
keV; Manzo et al. 1997).

Following the reduction procedures outlined by \scite{kong00}, we
applied an extraction radius of $4'$ centred on  the source position
for both LECS and MECS images so as to obtain the source spectra. The MECS 
background was extracted by using long archival
exposures on empty sky fields. 
For the LECS spectrum, we extracted the background from two
semi-annuli in the same field of view as the source (see Parmar et al.
1999 for the reduction procedure). Both the
extracted spectra were rebinned by a factor of 3 so as to accumulate at least
20 photons per
bin and to sample the spectral full-width at half-maximum resolution (Fiore
et al. 1999). A systematic
error of 1\% was added to both LECS and MECS spectra to take account
of the systematic uncertainties in the detector calibrations (Guainazzi et
al. 1998). 
Data were selected in the energy ranges 0.8--4.0 keV (LECS),
1.8--10.5 keV (MECS) and 15--220 keV (PDS) to ensure a better
instrumental calibration (Fiore et al. 1999). A normalization factor was
included for the LECS and PDS relative to the MECS in order to correct
for the NFI's flux intercalibration (see Fiore et al. 1999).

The source was only detected in the 1--10~keV range, and the broad-band (1--10 keV) 
spectrum of GX\,339--4 from the
LECS and MECS data is satisfactorily ($\chi^2_{\nu}=1.3$ for 33
degrees of freedom) fitted by a single power-law with photon
index of $1.78\pm0.24$ plus absorption. We fixed the $N_H$ at
$5.1\times 10^{21}$~cm$^{-2}$ \cite{kong00}. The absorbed flux in the
2--10~keV band is $0.69\times 10^{-12}$~erg cm$^{-2}$s$^{-1}$.

\section{Searching for X-ray variability on long timescales}
\begin{figure}
\begin{center}
{\rotatebox{-0}{\psfig{file=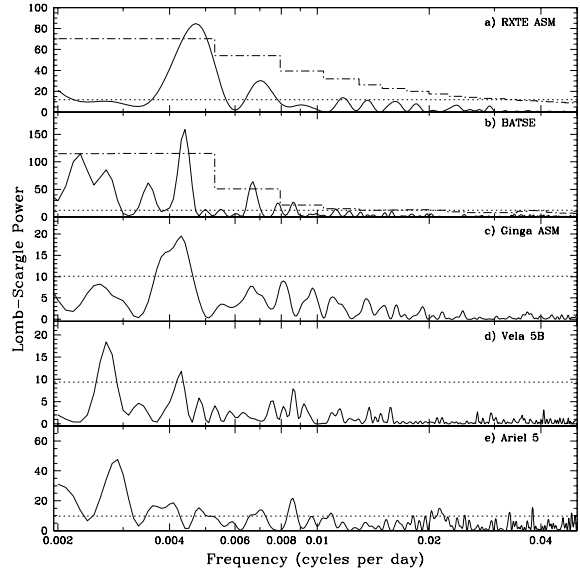,height=8.3cm,width=8.3cm}}}
\caption{Lomb-Scargle periodograms of GX\,339--4 as obtained by 
(a) \rxte\ ASM; (b) BATSE, (c) \ginga\ ASM, (d) {\it Vela 5B} and (e) {\it Ariel 5},
with significant peaks in the first four at 213, 227, 232 and 232d respectively.
The horizontal dotted line is the
99\% confidence level for white noise, while the dot-dash histogram is
the 99\% confidence level for red noise. The contribution of the red
noise component in the \ginga, {\it Vela 5} and {\it Ariel 5} data is small 
enough ($\lta$~1\%) to be ignored since the noise level is essentially limited 
by the white noise.}
\label{fig:combine_pow}
\end{center}
\end{figure}

From the light curves shown in the above section, GX\,339--4 underwent
several state transitions and showed quasi-periodic variability in the past 30
years. We therefore exploit the data collected from different
instruments to characterise its overall variability. 
In order to search for periodic phenomena, we used the Lomb-Scargle
periodogram (Lomb 1976; Scargle 1982; hereafter LSP),  
a modification of the discrete Fourier transform which is 
generalised to the case of uneven spacing. We have also employed the
Phase Dispersion Minimisation (PDM; \pcite{stellingwerf78}) to check
the results from the LSP, as the PDM is sensitive to non-sinusoidal
modulations. For the \rxte\ ASM data (see inset of Fig.~\ref{fig:batse}), we 
first
excluded the most recent HS and `off' state (after MJD 50815) from the
dataset as the flat-topped HS will dominate the LSP and the data
from the `off' state are non-detection. By applying the LSP, a broad peak was 
found at periods around 190--240~d, centered  at 213~d (see
Fig.~\ref{fig:combine_pow}a). We also plot in
Fig.~\ref{fig:combine_pow}a the 99\% significance level for both white
noise (Gaussian) and red noise. In determining the white-noise
confidence level, we generate Gaussian noise datasets with the same
time intervals and variance as the true data and then perform the LSP
analysis on the resulting datasets. The peak power in each periodogram
(which must be purely due to noise) was then recorded. This was
repeated 10,000 times for good statistics. However, the noise is not
necessarily Gaussian, it can also be frequency dependent with a
higher power at lower frequencies. By using the above method, strong
peaks at the low frequencies might give misleading results. In order
to  take into account such a noise contribution in the data, we
simulate the noise as a power law, which is also known as
red noise (e.g. \pcite{done92}). Quantitatively, the power spectrum of
the red-noise light curve will be given by $(1/\omega)^{\alpha}$,
where $\omega$ is frequency. Following an
implementation of the method of \scite{timmer95}, we generate
simulated light curves with the red-noise power spectrum, together with
the same time sampling and mean variance as the original data. Finally, we 
combine the white-noise simulated light curve as mentioned above, with the 
red-noise simulated light curve, by scaling the data points in each such that 
the relative contribution of both white and red-noise components is
matched with the real data. Following the method used previously to determine the white-noise 
confidence level, we derive the red-noise confidence level with the combined 
simulated light curve. Since red noise is frequency dependent, we need to 
compute the confidence level for a set of frequency bins. Hence it appears as 
a histogram rather than the continuous line for white-noise (see 
Fig.~\ref{fig:combine_pow}).

\begin{figure}
\begin{center}
{\rotatebox{-0}{\psfig{file=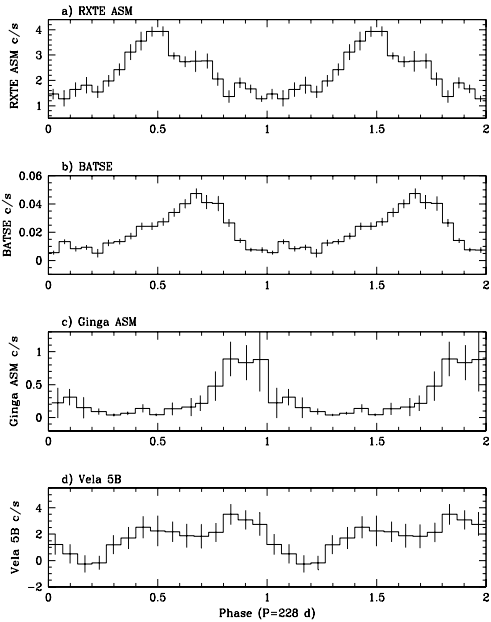,height=9cm,width=8.3cm}}}
\caption{Folded light curves of \rxte\ ASM (a); BATSE (b), {\it Ginga}
ASM (c) and
{\it Vela 5B} (d) on a period of 228~d which is based on a common
peak power determined in the LSP so that they are all on a common
ephemeris for comparison of the structure. Two cycles are shown for clarity. T$_0$ is 
set at the time of the first data point of {\it Vela 5B}.}
\label{fig:fold}
\end{center}
\end{figure}

The peak at 190--240~d is well above the 99\% confidence levels (as
determined by both methods) and we conclude that it is highly
significant.  The folded light curve of our {\it RXTE} ASM data on
228~d (which is consistent with the peaks in the \rxte, BATSE, \ginga\ and
{\it Vela 5B} periodograms, and such that they are all on a common
ephemeris for intercomparison of the structure so as to see whether
the variability is stable) is shown in Fig.~\ref{fig:fold}; T$_0$ is
defined by the first data point of {\it Vela 5B}. Analysis with the
PDM gave similar results to the LSP.

For the BATSE data, ten outbursts occurred during the past 7 years and
like the \rxte\ ASM data,
we exclude the recent high/soft and `off' state (after MJD~50800) in
calculating the LSP. The LSP shows a strong peak at $227\pm3$ days 
(Fig.~\ref{fig:combine_pow}b) which is
consistent with the {\it RXTE} ASM result.  The peak is very distinct
and exceeds the 99\% confidence level of both white and red
noise. The PDM analysis also confirmed this modulation. The folded
light curve of the BATSE data on 228~d is also shown in
Fig.~\ref{fig:fold}; T$_0$ is set at the time of the first data point
of {\it Vela 5B}. We also tried to exclude the HS data at around
MJD~48550 and 49450 (see \S\ref{sec:batse}) to calculate the LSP and
the result was indistinguishable from those presented above.  

The \ginga\ ASM monitored the source for 4.5 years and the four LS/HS/VHS 
state transitions on timescales of $\sim$~200--400 d are seen (see
Fig.~\ref{fig:ginga}). However, the sparse sampling prevents us
from calculating a good LSP. Nevertheless, we include all the data points
in searching for any variability. The strongest peak in the LSP is at
$\sim$~232~d, although it is broad and only marginally significant (see
Fig.\ref{fig:combine_pow}c). The folded
light curve of the {\it Ginga} data on 228~d is also shown in
Fig.~\ref{fig:fold}; T$_0$ is set at the time of the first data point
of {\it Vela 5B}.  

For the {\it Vela 5B} data, we used the whole dataset for calculating
the LSP. It should be noted that two HS occurred during these
observations (see Appendix; \pcite{markert73}) and hence the results should be
interpreted with great caution. The strongest peak in the periodogram of the 
{\it Vela 5B} data is at $\sim$~365 days and is caused by an annual variation 
in the spacecraft environment \cite{priedhorsky83}. The
LSP reveals a second peak at 232~d which is just above the 99\% confidence 
level (see Fig.~\ref{fig:combine_pow}d).
Although the sensitivity of {\it Vela 5B} is poorer than current X-ray
satellites, a clear variability is seen in the light curve (Fig.~\ref{fig:av}) 
and the long timebase (over 10 years) increases the sensitivity to these 
timescales.
 This variability was also detected in the PDM analysis.
The folded light curve on 228 days is shown in Fig.~\ref{fig:fold};
T$_0$ is again set at the time of the first data point of {\it Vela 5B}.

The {\it Ariel 5} ASM data do not show the variation as clearly as  
the {\it RXTE} ASM data, but some variability can be discerned.  LS/HS
transitions might occur during the {\it Ariel 5} observations but
there is no pointed observations to confirm the `state' of the
source. We here used all the data in calculating the LSP. The LSP 
(Fig.~\ref{fig:combine_pow}e) shows a peak near 365 days, again due to the 
annual variation in the count rate from solar X-ray scattering or fluorescing  
from the Earth's atmosphere \cite{priedhorsky83}.

\section{Discussion}
\subsection{Long-term X-ray variability}
Through the extensive dataset obtained with the \cgro/BATSE and
\rxte/ASM, we find evidence of variability on timescales
$\sim$~190--240~d in GX\,339--4 (it should be noted that the recent HS
and `off' state data are excluded in both datasets and at least one HS
occurred in the BATSE data). Observations
from \ginga\ and {\it Vela 5B} also suggest a similar timescale but the data
consist of state transitions. Such a variability has been
noted previously by \scite{nowak99} for the \rxte\ ASM data. They 
also pointed out that the
observed long-term X-ray variability is not a strict clocking
phenomenon but is more likely to be a characteristic timescale. From
our analysis, this timescale is also manifested in the BATSE, \ginga\
and {\it Vela 5B} data.  

In addition, the broad peak in the LSP of the \rxte\ ASM data (see
Fig.~\ref{fig:combine_pow}) also hints that such variability is not
strictly periodic and can only be regarded as a characteristic
timescale. From the folded light curves of different datasets 
(Fig.~\ref{fig:fold}), the
peaks occur almost in anti-phase, suggesting the aperiodic nature.
 Indeed, such 
quasi-periodic long-term variability (usually referred to as the {\it 
super-orbital} period) has also been seen in other X-ray binaries such as 
Cyg\,X--2, LMC\,X--3 and Cyg\,X--1 (see e.g. \pcite{kong98,paul00,kitamoto00}),
 but no clear understanding of its origin has yet been established (see 
\pcite{wijers99}).
As noted by \scite{nowak99}, the observed long-term X-ray variability
of GX\,339--4 could be due to a combination of a quasi-steadily
precessing disc at large radii with coronal structure changes on small
radii. If this is the case, X-ray irradiation can give rise to a
precessing disc \cite{dubus99} and it might even explain the LS/HS
transition of the source. Actually, results from the \ginga\ and {\it
Vela 5B} data suggest that the LS/HS transitions and the outburst-like events 
during the low/hard
state (as seen in the \rxte\ and BATSE data)
may be due to the same mechanism, although the data is not
well-sampled. Therefore, it is possible that such quasi-periodic outbursts 
might
actually be a characteristic precursor to the LS/HS transition.
In the next subsection, we will discuss the role of
X-ray irradiation in the state transition and also the origin of the
long-term X-ray variability. 

Our \sax\ observations presented here indicate that the source was
fainter than the previous observation in 1999 August (see
\pcite{kong00}) by a factor 3 in half year; therefore the source is
still in the `off' state. The energy spectra of both observations
are very similar to those obtained in the LS. Hence, there is no doubt
that the `off' state is an extended LS (see discussions by
\pcite{kong00,corbel00}) and we may now refine the source as a transient
rather than a persistent source. In fact the luminosity during the
`off' state is in the upper end of the quiescent luminosity of typical
X-ray transients (see e.g. \pcite{kong00}). The results also further
confirm that during quiescence, significant X-ray variability can be
seen. Such variability in quiescence is also found in several black hole
transients such as GS\,2023+338 (\pcite{wagner94}; \pcite{kong01}), 4U\,1630--47 \cite{parmar97} and A0620--00 
\cite{asai98,menou99}.

\subsection{LS/HS transition in 1998}
As noted in \S\ref{sec:339rxte} and \ref{sec:batse}, GX\,339--4 underwent a 
LS/HS transition in 1998 (see Fig.~\ref{fig:batse}; also
\pcite{belloni99}) which differs somewhat from the previous ones observed with the \ginga\ ASM.
After 
the source reached the peak of the HS, it spent nearly 200
days at a constant flux level of $\sim$~20 \rxte/ASM cts/s. 
Such a flat-topped X-ray light curve for GX\,339--4 is different
from the `standard' fast-rise exponential decay profile, as observed
from various X-ray transients (see \pcite{chen97} for various examples
of different outburst patterns). The state transition mechanism for such a
persistent source is still a puzzle and observations in different
states suggested that it is most likely associated with the mass
accretion rate (e.g. \pcite{mendez97}). A change in the mass accretion
rate can be caused by a thermal instability in the accretion disc
(disc instability model, or DIM) and in fact it is widely accepted
that the outbursts in BHSXTs are due to this mechanism
(e.g. \pcite{esin00}). Interestingly, the DIM was originally developed
to describe the outbursts in dwarf
novae (DN; see e.g. \pcite{cannizzo93} for a review) which contain a white dwarf
as the compact object; it was suggested quite early that it also
applies in BHSXTs (e.g. \pcite{paradijs84,huang89,mineshige89}). In LMXBs, X-ray irradiation
can influence the stability of the accretion disc \cite{paradijs96}
and thus a DIM which includes X-ray irradiation was developed
\cite{dubus99}. It is worth noting that the recent HS
spectroscopic observations of GX\,339--4 obtained by \scite{wu00} also 
indicate that
the accretion disc is heated by soft X-ray irradiation, suggesting
that irradiation plays a role in the state transition. Irradiation of the 
outer disc by X-rays from the inner disc gives a stability criterion, from 
which the minimum mass accretion rate for a stable, steady-state accretion in 
an X-ray irradiated disc can be expressed as \cite{dubus99}:

\begin{eqnarray}
\mdot_{crit}^{irr} \approx 2.0 \times 10^{15}\left(\frac{M_1}{\msun}\right)^{0.
5}\left(\frac{M_2}{\msun}\right)^{-0.2} \times \nonumber \\
P_{hr}^{1.4} 
\left(\frac{\mathcal{C}}{5\times10^{-4}}\right)^{-0.5} \mbox{g s$^{-1}$,}
\label{eq:irr}
\end{eqnarray}

where $M_1$ is the black hole mass, $M_2$ is the mass of the companion, 
$P_{orb}$ the orbital period in hours. $\mathcal{C}$ is a parameter to 
describe the properties of irradiation through:

\begin{equation}
T^4_{irr}=\mathcal{C}\frac{\mdot c^2}{4\pi\sigma R^2},
\end{equation}

where $T_{irr}$ is the irradiation temperature and $\sigma$ is the
Stefan-Boltzmann constant. 
For a stable X-ray irradiated disc during the HS, $\mdot_{crit}^{irr}$
can be estimated from the HS observations by 
\scite{belloni99}, in which they have fitted the spectrum with a power law 
plus a multi-colour disc blackbody model. By using the derived inner disc 
radius, $R_{in}\sqrt{\cos i}$ of $25.4 D_{4}$~km (where $D_4$ is the distance 
to the source in units of 4~kpc; \pcite{zdziarski98}) and the blackbody 
temperature, $T_{in}$ of 0.72~keV, we can calculate the bolometric luminosity 
of the disc blackbody component as

\begin{eqnarray}
L_X &=& 4\pi R_{in}^2 \sigma T_{in}^{4} \nonumber \\
    &=& 2.2\times 10^{37} D_4^2/\cos i \quad\mbox{erg s$^{-1}$}.
\end{eqnarray}    

The accretion rate can be estimated via the relation $L_X=0.5
GM_1\mdot /R_{in}$, which leads to $\mdot\approx 1.7\times 10^{17}
D_4^2$~g s$^{-1}$ if we assume an orbital inclination of about
$15^{\circ}$ \cite{wu00} for a 14.8-hr orbital period
\cite{callanan92}. 
Assuming an estimated mass of the black hole of 
5\msun\ \cite{zdziarski98} and companion star of 0.4\msun\ \cite{callanan92}, 
we obtain $\mathcal{C}=9.4\times10^{-4}$ from Eqn. 1, which is
slightly higher than 
$5\times10^{-4}$ obtained by comparison with other results in the
literature (see \pcite{dubus99}). According to the scenario outlined
above, the mass of the compact object is limited to be $\sim$~3--5\msun.

From the
previous LS observations, the estimated mass accretion rate is about
$5\times 10^{16}$~g s$^{-1}$ (e.g. \pcite{wilms99,belloni99}) which is
a factor of 3 
below the critical accretion rate for a stable disc and hence the disc
may be subject to the DIM in order to give rise to a LS/HS transition. Since
the accretion rate during the LS is very close to the critical value,
a slight increase in the accretion rate by a small factor would
produce a stable disc.  As a result,  a shorter interval to the
next state transition would be expected. Unlike the SXTs in which the
outburst timescale is normally longer than ten years, GX\,339--4 has
much more frequent state transitions, as seen in e.g. the \ginga\ ASM
data (see Fig.~\ref{fig:ginga}). In fact, the 232-d variability seen
in the \ginga\ and {\it Vela 5B} data has already hinted at such behaviour. 

The most interesting behaviour of the 1998 HS is that the observed flat-topped 
X-ray light curve of GX\,339--4 bears
characteristics of Z Cam-type DN in which standstills (i.e. the
brightness remains constant) occasionally interrupt the recurrent
outbursts. Such a scenario can be explained
by the DIM \cite{meyer83,king98} in which the accretion rate before
standstill is very close to a critical value and a sudden
increase in the accretion rate (e.g. by a starspot or irradiation induced
mass overflow) triggers the standstill state. The light curve of
GX\,339--4 is even more similar to the model proposed by
\scite{king98} in which the standstill is at a higher luminosity than
the normal outbursts. For Z Cam systems, the intensity of the standstills is
lower than the maximum of the outbursts (\pcite{meyer83}), but we
should note that the models for explaining the Z
Cam systems are used to account for optical light curves. 

The second similarity
between GX\,339--4 and Z Cam systems concerns the recurrence of
outbursts between standstills. From our earlier period analysis of the
archival GX\,339--4 data over 30 years, we find a variability timescale of  $\sim$190--240~d 
and we
suggest this to be the recurrent outbursting behaviour seen in Z Cam
systems. As
pointed out by Dubus et~al. (in preparation), the accretion rate can
vary on timescales of 10--100 days when a viscously unstable disc is
irradiated by a constant X-ray flux. If irradiation is crucial for
determining the accretion rate of GX\,339--4, this may explain the variability
of the `mini'-outbursts observed during the LS. There is also an
indication that the accretion rate of GX\,339--4 was building up
before the HS in 1998
as the average intensity level of both \rxte\ ASM and BATSE data has a
slight increasing trend especially after MJD~50000 (see
Fig.~\ref{fig:batse}). In addition, the HS occurring at MJD~48550 and
49450 (see \S\ref{sec:batse}) were just after a strong X-ray outburst in the LS
(see Fig.~\ref{fig:batse}). Perhaps this takes the accretion rate in the
LS to be close to the critical value and finally triggers the state
transition. As a result, this HS may be simply a strong, prolonged
outburst in an otherwise LS. In fact, from the period analysis of the {\it Vela
5B} and \ginga\ data for which state transitions are included in the
calculation, the timescale of variability resembles that found in the 
\rxte\ ASM
data in the LS. Perhaps the LS outbursts may be excursions
towards the HS but which they do not quite reach.
If this is the case, we would expect similar
spectral evolution during the LS outburst, just like during a state
transition. However, such outbursts are usually short and therefore 
observations
are difficult to arrange. The most likely observations available are from
\scite{wilms99}, in which an \rxte\ pointed observation of GX\,339--4 was made near
the outburst (MJD~50710) just before the LS/HS transition. Although
the energy spectrum appeared to be slightly softer, it is still
consistent with other LS observations within the uncertainties. It is worth
noting that Cyg\,X--1 also shows LS/HS transitions occasionally and it
has flaring activities during the LS. Moreover, the long HS of
Cyg\,X--1 between
1996 May and August resembles that seen in GX\,339--4, but with
more flares during the HS \cite{zhang97}. Therefore, it is possible that the
state transition of Cyg\,X--1 is due to a similar mechanism to that
discussed here (see however \pcite{zhang97,esin98}). More recently,
LMC\,X--3 was also shown to have recurring state transitions but
the driving force is very likely to be due to other mechanisms \cite{boyd00,wilms00}. 
Given 
that the accretion rate in the LS is close to the critical value, similar LS/HS 
transitions would be expected in the near future and multi-wavelength 
observations of such transitions, particularly the correlation between the 
X-ray, optical and radio emissions will constrain the accretion disc structure 
and behaviour of the secondary star.

\section*{Acknowledgments}
We are grateful to Colleen Wilson-Hodge and Ken Watanabe for providing
the updated BATSE data. We also thank Christine Done for the
red-noise generator code. AKHK is supported by a Hong Kong Oxford
Scholarship. This paper utilizies quick-look results provided by the
ASM/RXTE team and data obtained through the HEASARC Online Services of
NASA/GSFC. 

\appendix
\section{Previous pointed X-ray observations of GX\,339--4}

GX\,339--4 has been observed by several X-ray satellites in the past
30 years and the results from those pointed observations are crucial
to determine the `state' of the source. We compile here a list of
pointed observations of GX\,339--4 from the literature (see Table A1). Note 
that the
`state' quoted is determined by spectral and temporal
analysis. Although the X-ray flux level can more or less
reflect the `state' of the source, on several occasions the X-ray flux in the 
LS is actually higher
than the HS (e.g. 1981 May and 1984 May). Hence the X-ray intensity
itself is not an accurate indicator of X-ray `state'.

\begin{table*}
\small
\caption{Previous X-ray observations of GX\,339--4.}
\begin{center}
\begin{tabular}{l c c c c}
\hline
Date & X-ray state & Flux (mCrab) & Satellite &References\\
\hline
1971 October to 1972 January & LS & $\sim$~10--40 (1--6~keV)& {\it OSO-7}&1\\
1972 February & HS & $\sim$~90 (1--6~keV)& {\it OSO-7}& 1\\
1972 May & LS/off& $<$~5 (1--6~keV) & {\it OSO-7}& 1\\
1972 December & HS & $\sim$~300 (1--6~keV)& {\it OSO-7} &1\\
1981 March & LS/off &$<$~30 & {\it Hakucho}&2\\
1981 May& LS & 160 (0.1--20~keV)& {\it Hakucho}&2, 3\\
1981 June& LS to HS &$\sim$~160--600 (3--6~keV) & {\it Hakucho}& 2\\  
1982 May & LS/off &$<$~15 & {\it Hakucho} & 2\\
1983 May & HS &300 (2--10~keV) & {\it Tenma}& 4\\
1984 March \& May & IS & 90 (2--10~keV) & {\it EXOSAT}& 5 \\ 
1984 May & HS & 120 (0.1--20~keV)& {\it EXOSAT} &3\\
1985 April & LS/off & 1.7 (0.1--20~keV) & {\it EXOSAT}& 3\\
1987 June & LS/off &$\sim$~13 (1--6~keV) & {\it Ginga}& 6, 7\\
1987 July & LS/off &$\sim$~26 (1--6~keV) & {\it Ginga}& 6, 7\\
1988 September & VHS & 900 (1--6~keV) & {\it Ginga}& 7, 8\\
1989 August & LS/off& $\sim$~12 (1--6~keV) & {\it Ginga}& 7, 9\\
1990 April & LS& 100 (3--10~keV) & {\it Granat}& 10\\
1990 August & HS &100--230 (3--10~keV) & {\it Granat}& 10\\
1991 February & off &$<$~6 (3--10~keV) & {\it Granat}& 10\\
1991 August to October & LS to HS &$\sim$~100--300 (3--10~keV) &{\it
Granat}& 10\\
1992 February & off &$<$~17 (3--10~keV) & {\it Granat}& 10\\
1993 September & off &$<$~0.01 (2--10~keV) & {\it ASCA} & 11\\
1994 August & LS &$\sim$~30 (3--9~keV) & {\it ASCA}& 12\\
1995 September & LS &50 (3--9~keV) & {\it ASCA}& 12\\
1996 July & LS & 70 (2--12~keV) & \rxte & 13\\
1997 February to October & LS & $\sim$~70 (2--12~keV)& \rxte& 12\\
1998 January to February & HS & 160--270 (2--12~keV)& \rxte& 13\\ 
1998 August& HS & 260 (2--12~keV) & \rxte &14\\
1999 August & off &0.1 (2--10~keV) & \sax &14, 15\\
2000 March  & off &0.03 (2--10~keV) & \sax& this work\\
\hline
\\
\multicolumn{4}{l}{References: (1) Markert \etal 1973; (2) Motch \etal
1985; (3) Ilovaisky \etal 1986; (4) Makishima \etal 1986;}\\ 
 \multicolumn{4}{l}{(5) M\'endez \& van der Klis 1997; (6) Callanan
\etal 1992; (7) Kitamoto \etal 1992; (8) Miyamoto \etal 1991;}\\
\multicolumn{4}{l}{(9) Steiman-Cameron \etal 1990; (10) Grebenev \etal
1993; (11) Asai \etal 1998; (12) Wilms \etal 1999;}\\
\multicolumn{4}{l}{(13) Belloni \etal 1999; (14) Corbel \etal 2000; (15) Kong 
\etal 2000}
\end{tabular}
\end{center}
\label{tab:previouslog}
\end{table*}

\end{document}